

\let\oldeqno = \eqno
\def\eqno#1{\oldeqno{\hbox{#1}}}

\def\pmb#1{\setbox0=\hbox{$#1$}%
\kern-.025em\copy0\kern-\wd0
\kern.05em\copy0\kern-\wd0
\kern-.025em\raise.0433em\box0}
\def\tr{\,{\rm tr}\,}

\def\pp{\partial}

\def\TC{\Theta\kern-.1em\lower0.6ex\hbox{${}_{C}$}}
\def\br{{\bf r}}
\def\bp{{\bf p}}
\def\bq{{\bf q}}
\def\brp{{\bf r'}}
\def\N1{k}
\def\M1{l}
\magnification=1200

\hsize=6.7truein
\hoffset=-.1in
\vsize=9.0truein
\voffset=-.1in

\tolerance 600
\font\eightrm = cmr8
\def\footnoterule{\kern-3pt \hrule width \hsize \kern2.6pt}
\pageno=0 \footline={\ifnum\pageno>0 \hss --\folio-- \hss \else\fi}
\baselineskip 14pt plus .5pt minus .5pt
\parskip=2pt

\centerline{\bf Non--Abelian Chern--Simons Particles and their
Quantization\footnote{*}
{\baselineskip 11pt \eightrm
This work is supported in part by funds provided by
the U.~S.~Department of Energy (D.O.E.) under contracts
\#DE-AC02-76ER03069 and \#DE-AC02-89ER40509, as well as in part by the National
Science Foundation under contracts \#INT-910559 and \#INT-910653}}
\smallskip
\centerline{{\bf}}
\vskip 24pt
\centerline{Dongsu Bak and R. Jackiw}
\vskip 12pt
\centerline{\it
Center for Theoretical Physics,
Laboratory for Nuclear Science,
and Department of Physics}
\centerline{\it
Massachusetts Institute of Technology,
Cambridge, Massachusetts~~02139~~~U.S.A.}
\vskip 24pt
\centerline{So-Young Pi}
\vskip 12pt
\centerline{\it
Department of Physics,
Boston University, Boston, Massachusetts~~02215~~~U.S.A.}
\vfill
\midinsert
\narrower
\medskip
{\baselineskip 14pt plus .5pt minus .5pt
\centerline{\bf ABSTRACT}
A many--body Schr\"odinger equation for non--Abelian Chern--Simons particles is
obtained
from both point--particle and field--theoretic pictures. We present
a particle Lagrangian and a field theoretic Lagrange density,
and discuss their properties. Both are quantized by the symplectic method
of Hamiltonian reduction. An $N$--body Schr\"odinger equation for the particles
is obtained from both starting points.
It is shown that
the resulting interaction between particles can be replaced by non--trivial
boundary
conditions. Also, the equation is compared with the one given in the
literature.
    \par}
\endinsert
\vfill
\centerline{
Submitted to:
{\it Physical Review D }}
\vfill
\line{CTP \# 2276 \hfil
hep-th/9402057\hfil
January 1994}
\eject
\goodbreak\bigskip
\line{\bf I.~~Introduction \hfil}
\nobreak\medskip\nobreak\noindent

Systems of point particles carrying  non--Abelian charge in interaction
with a  non--Abelian gauge field have been under investigation for over two
decades.$^{1}$
Recently, such a model in 2+1 dimensions with Chern--Simons kinetic action has
gained
attention owing to its peculiar long range interactions.$^{2}$

In general, particles interacting via the Abelian
Chern--Simons term acquire fractional statistics,$^{3}$ and are
called anyons.$^4$ Anyons play a role
in the fractional quantum Hall effect$^5$ and perhaps also in
high temperature superconductivity.$^{6}$

A system of non--relativistic point particles with Abelian Chern--Simons
interactions has been previously discussed in [7] and [8]. Non--relativistic
quantum dynamics can be equivalently described by a point particle Lagrangian
or by a field theory; upon quantization both lead to an $N$--body Schr\"odinger
equation, with the Aharonov--Bohm (Ehrenberg--Siday) interaction.
These two approaches are pursued
respectively in Refs.~[7] and [8]. Moreover when the field theoretical
description is
analyzed classically, self--dual solitons are found.$^{9}$
(Soliton solutions exist when a  quartic self--interaction
is included. Quantum mechanically this corresponds to a $\delta$--function
hard--core interparticle potential, whose only effect is to regularize the
Chern--Simons
interaction, protecting its classical conformal invariance$^{10}$against
quantum
anomalies.$^{11}$)

In this paper we extend our previous investigations to the non--Abelian case.
Much
has already been established on this problem. Non--Abelian self--dual
solitons have been found
in the corresponding non--relativistic field theory,$^{12}$which however has
never
been quantized. The $N$--body Schr\"odinger
equation has been posited on the basis of non--Abelian  braid group
investigations,$^{13}$ but without a derivation from first principles.
A systematic derivation from a particle Lagrangian for the $SU(n)$ group was
given
subsequently.$^{14}$ In these works use is made of the
Knizhnik--Zamolodchikov connection$^{15}$---a non--Hermitian choice of gauge,
which requires a  non--trivial compensating measure in the normalization
integral.

In Section~II, we reconsider the derivation of the Schr\"odinger equation
from a particle Lagrangian. We use the symplectic methods of Hamiltonian
reduction$^{16}$ and can accommodate any non--Abelian gauge group, provided it
is equipped with an invariant metric. Furthermore, we choose
a real gauge and dispense with
the complex Knizhnik--Zamolodchikov connection. Both distinguishable and
identical particles are considered.

In Section~III, we follow the alternative route to the
Schr\"odinger equation, by beginning with a field theoretic Lagrange density.
Again using the symplectic method and constructing the $N$--particle state
as  an eigenstate of the number operator, allows deriving the
$N$--body equation, which coincides with the one obtained in Section~II for
identical particles.

In Section~IV, the Schr\"odinger equation is analyzed. We show that the
interaction
between the particles may be replaced by a boundary condition and is equivalent
to the
one given by the Knizhnik--Zamolodchikov connection.

An Appendix is devoted to a symplectic quantization of an arbitrary Lie
algebra, possessing an invariant metric. This is accomplished without explicit
passage to Darboux (canonical) variables.

\goodbreak\bigskip
\line{\bf II.~~Non--Abelian Chern--Simons particles for an arbitrary group
\hfil}
\nobreak\medskip\nobreak\noindent

The non--Abelian charge carried by a point particle may be viewed as a
classical internal
degree of freedom that produces a non-Abelian gauge field. Upon quantizing
this degree of
freedom, the charge operator acquires a spectrum, which leads to
a multiplet
structure
of the particle state.

Let us begin with a classical description on the level of equations of motions.
Since
the source produces a  non--Abelian gauge field, we expect that a field
equation
relates the field to the source.
$$\eqalign{
&(D_\mu F^{\mu\nu})^a + {\kappa \over 2} \epsilon^{\nu\alpha\beta}
(F_{\alpha\beta})^a =  j^{\nu} I^a\cr
&D_\mu\equiv \pp_\mu+[A_\mu,\cdots]\cr}\eqno{(1)}
$$
A Chern--Simons term with strength $\kappa$, is included because we are
considering
a (2+1)-dimensional theory.
Here $j^0(t,\br)=\delta \left({\bf r}-{\bf q}(t)\right)$ and
${\bf j}(t,\br)={\dot {\bf q}}(t)\delta ({\bf r}-{\bf q}(t))$ comprise the
conserved particle
current density, $j^\mu$, for a particle located in the plane at ${\bf q}(t)$
and
$I^a(t)$ is the non--Abelian charge. [Henceforth, we supress a common time
argument in all
dynamical quantities.]
Since the left side of (1) is covariantly conserved, so must be the right side.
This
puts an additional condition on $I$:
$$\eqalign{
D_\mu(j^\mu I)&= (\pp_\mu j^\mu) I + j^\mu D_\mu I\cr
 &= \delta \left({\bf r}-{\bf q}\right)D_q I=0\cr}
\eqno{(2)}
$$
with $D_q\equiv \pp_t +[A_q, \cdots]$,  $A_q\equiv A_0 -
{\dot {\bf q}}\cdot {\bf A}$, $A_\mu \equiv A_\mu^a T_a$, $I\equiv I^a T_a$,
and the generators
$T_a$ (in
an arbitrary representation) satisfy $[T_a,T_b]=f_{ab}\,^c T_c$. Our
space--time
metric tensor is ${\rm diag}(1,-1,-1)$.
The force
law, a  non--Abelian generalization of the Lorentz force, is given  by
$$
M{\ddot {\bf q}} =I_a ({\bf E} +{\dot {\bf q}}\times { B})^a \eqno{(3)}
$$
where the non-Abelian electric and magnetic fields are, respectively,
$E^{ia}=F^{ioa}$,
$\epsilon^{ij}B^{a}=-F^{ija}$ and $({\dot {\bf q}}\times { B}^a)^i\equiv
\epsilon^{ij}
{\dot q}^j B^a$.
Eqs. (1), (2) and (3) are the celebrated Wong equations.$^1$

Notice that
presentation of the Lorentz force equation (3) requires lowering
the group index on the non--Abelian charge. For semi--simple groups
this is accomplished with the non--singular Cartan--Killing metric $g_{aa'}=
-{1\over 2}f_{ab}\,^cf_{a'c}\,^b$. For non--semi--simple groups, the
Cartan--Killing
metric is singular, so we must assume that there exists another
invariant non--singular form on the Lie algebra which can serve as a metric.
This is  the only assumption we need to make about the structure of the
gauge group.

Under a local gauge transformation by a group element $U$, the gauge potential
$A$ and
the particle coordinate ${\bf q}$ transform as
$$
A_\mu \rightarrow U^{-1}A_{\mu} U + U^{-1}\pp_{\mu} U,\ \ \ \ \ \ \
 {\bf q} \rightarrow {\bf q}
\eqno{(4)}
$$
To preserve gauge covariance of the equations of motion, the charge $I$ must be
transform
covariantly.
$$I\rightarrow U^{-\! 1}I\, U
\eqno{(5)}
$$

Since we are interested only in the Chern--Simons kinetic term, we drop
the non--Abelian Maxwell (Yang--Mills) term in (1).
Then the  equations for the $N$--particle system read
$$
{\kappa \over 2} \epsilon^{\nu\alpha\beta}
(F_{\alpha\beta})^a = \sum_{k=1}^N
 j_k^{\nu} I_k^a
\eqno{(6.a)}
$$
$$\dot I_{\N1}^a + f_{bc}\,^a[A_0 ({\bf q}_\N1)-
{\dot {\bf q}}\cdot {\bf A} ({\bf q}_\N1)]^b I_{\N1}^c = 0
\eqno{(6.b)}
$$
$$
M_k{{\ddot {\bf q}}_{k}} =I_{ka} [{\bf E}({\bf q}_\N1) +
{\dot {\bf q}_k}\times {B}({\bf q}_\N1)]^a
\eqno{(6.c)}
$$
Eqs.~(6.a-c) are the Euler--Lagrange equations for the Lagrangian
$$\eqalign{L&= {1\over 2}{\textstyle \sum}M_\N1 {\dot \bq}^2_\N1 +2{\textstyle
\sum}
\tr K_\N1 g_\N1^{\! - 1}{\dot g}_\N1  +2 \int d^2\! r {\textstyle \sum}
 \tr j_\N1^{\nu} I_\N1 A_\nu ({\bf q}_\N1)\cr &-\kappa \int d^2\! r
\epsilon^{\alpha
\beta\gamma}
\tr (
A_\alpha \pp_\beta A_\gamma +{2\over 3}A_\alpha A_\beta A_\gamma) \cr}
\eqno{(7)}
$$
Here ``$\tr$'' stands for the trace, with $T_a$ normalized by
$\tr T_aT_b =-{1\over 2}g_{ab}$; when the trace is not available,
it is replaced by a non--degenerate bilinear form on the Lie algebra, which
also defines the metric. The  $K_\N1$ are time--independent elements (in
$N$ copies) of the Lie algebra (one for each $k$) and $I_k$ is defined as
$$I_\N1 = g_\N1 K_\N1 g_\N1^{\! - 1}
\eqno{(8)}
$$
where $g_\N1$  are time--dependent group elements, responding to
a gauge transformation by $ g_\N1^{\ } \rightarrow g'_\N1=U^{-1}g_\N1^{\ }$;
as a consequence the middle two terms in (7) combine to
$2{\textstyle \sum}
\tr K_\N1 g_\N1^{\! - 1}(\pp_t+A_q)g_\N1$.
The Euler--Lagrange equations (6.a-c) are derived by varying $A_\mu$, $g_k$ and
 ${\bf q}_k$.

Upon introducing canonical momenta for the positions ${\bf q}_\N1$,
$${\bf p}_\N1 \equiv {\pp L\over\pp{\dot {\bf q}}_\N1} =M_\N1 {\dot{\bf q}}_\N1
+  I_{\N1 a}{\bf A}^{\! a}({\bf q}_\N1)  \eqno{(10)}
$$
and by performing a Legendre transformation, the Lagrangian is
converted into a first--order
expression.
$$\eqalign{L&= {\textstyle \sum}[\bp_\N1\cdot{\dot \bq}_\N1 +
2\tr K_\N1 g_\N1^{\! - 1}{\dot g}_\N1^{\ } ]+ \kappa\int d^2\! r {\dot A}_{1a}
A_{2}^a\cr
&-{\textstyle \sum}{1\over 2M_\N1} (\bp_\N1
-I_{\N1 a}{\bf A}^{\! a}({\bf q}_\N1) )^2
 \cr
 &- \int d^2\! r \left\{A_0^a(\kappa B_a +\rho_a)
+{\kappa\over
2}[\pp_t(A_1^aA_{2a})+\pp_1(A_2^aA_{0a})-\pp_2(A_1^aA_{0a})]\right\}\cr
&\rho^a \equiv {\textstyle \sum}I_{\N1 }^a
\delta \left({\bf r}-{\bf q_\N1}\right)\cr}
\eqno{(11)}
$$
Under a gauge transformation  the dynamical variable ${\bf p}_k$ transforms as
$$
\bp_\N1 \rightarrow
\bp_\N1+2\tr I_\N1\nabla U({\bf q}_\N1^{\ }) U^{-1}({\bf q}_\N1^{\ })
\eqno{(12)}
$$
and the transformed Lagrangian  differs by total derivative
$$L\rightarrow L+\kappa \int d^2\! r \epsilon^{\alpha\beta\gamma}[ \pp_\alpha
\tr (
\pp_\beta U
U^{-1} A_\gamma)+{1\over 3}
\tr U^{-1} \pp_\alpha UU^{-1} \pp_\beta UU^{-1} \pp_
\gamma U]\eqno{(13)}
$$
The gauge invariance of the quantum theory requires a
quantized coupling constant $4\pi\kappa\in {\bf Z}$, when the gauge group is
compact and non--Abelian.$^{17}$

To quantize the system, we follow symplectic reduction method.$^{16}$ Note
that the Lagrangian in (11) is constrained by Lagrange multipliers $A^a_0$ and
the
corresponding constraints are
$$B^a+ {1\over\kappa }\rho^a = 0
\eqno{(14)}
$$
We solve the constraints by expressing $A_2$ in term of other quantities.
$$
A_2^a({\bf r}) = \int d^2\! r'{\cal M}^a\,_{b}({\bf r},{\bf r'})[\pp'_2
A_1^b({\bf r'})+
{1\over \kappa}\rho^b({\bf r'})]
\eqno{(15)}
$$
where ${\cal M}^a\,_{b}({\bf r},{\bf r'})$ is defined through the following
relations,
$$\eqalign{
 &\int d^2\! r' {{\cal M}^{-1}}^a\,_{b}({\bf r},{\bf r'})c^b({\bf
r'})=\pp_1c^a({\bf r})+
f_{bc}\,^a A_1^b({\bf r})c^c({\bf r})\cr
& \int d^2\! r' {\cal M}^{a}\,_b({\bf r},{\bf r'}){{\cal M}^{-1}}^b\,_{c}
({\bf r'},{\bf r''})=
\delta({\bf r}-{\bf r''}) \delta^a_c \cr}
\eqno{(16)}
$$
Inserting (15) into (11), we are left with
$$\eqalign{L&= \sum [\bp_\N1\cdot {\dot \bq}_\N1\! +\!
2\tr K_\N1 g_\N1^{\! - 1}{\dot g}_\N1^{\ } ]+ \kappa\int d^2 r  {\dot
A}_{1a}(\br)\!
\int d^2\! r' {\cal M}^a\,_{b}({\bf r},{\bf r'})[\pp'_2 A_1^b({\bf r'})\!+\!
{1\over \kappa}\rho^b({\bf r'})]\cr
&-\sum {1\over 2M_\N1} \left[(p^x_\N1
\!+\!I_{\N1 a}A^a_1({\bf q}_\N1) )^2  +(p^y_\N1
\!+\! I_{\N1 a}\!\!\int d^2\! r'{\cal M}^a\,_{b}
({\bf q}_k,{\bf r'})[\pp'_2 A_1^b({\bf r'})\!+\!
{1\over \kappa}\rho^b({\bf r'})])^2\right]
 \cr}
\eqno{(17)}
$$
where we ignore irrelevant total derivative terms. While the constraints have
been
eliminated, the symplectic one--form in the Lagrangian (17) is still not of a
desired canonical form. To achieve that, let us make a Darboux
transformation. We introduce new (primed) variables with the relations,
$$\eqalign{
 g_\N1 &= Ug'_\N1\cr
{\bf A} &= U{\bf A\!'} U^{-1} - U\nabla U^{-1}\cr
\bp_\N1 &= {\bp'}_{\!\! \N1}-2\tr I_\N1 \nabla U U^{-1}
\cr}
\eqno{(18)}
$$
where $A_2$ is given by (15) and $U$ will be specified presently.
The Lagrangian (17) becomes
$$\eqalign{L&= {\textstyle \sum}[{\bp'}_{\!\!\N1}\cdot{\dot \bq}_\N1 +
2\tr K_\N1 {g'}_{\! \N1}^{-1}{\dot {g}}'_\N1 ]+ \kappa\int d^2\! r {{\dot
A}'}_{1a}
{A'}^a_{2} \cr
&-{\textstyle \sum}{1\over 2M_\N1} \left[{\bp'}_{\!\!\N1}
-I'_{\N1 a}{{\bf A}\!'}^{a}({\bf q}_\N1) \right]^2
-{\kappa\over 3}\int d^2\! r \epsilon^{\alpha\beta\gamma}
\tr U^{-1} \pp_\alpha UU^{-1} \pp_\beta UU^{-1} \pp_
\gamma U \cr}
\eqno{(19)}
$$
where irrelevant total derivative terms are again ignored.

In the Lagrangian (19), $A'_2$
may be given in two different ways. First, using (15) and (18), $A'_2$ is
$$
A'_2= U^{-1}\int\!\! d^2\! r'T_a{\cal M}^a\,_{b}
({\bf r},{\bf r'})[\pp_2 A_1^b({\bf r'})\! +\!
{1\over \kappa}\rho^b({\bf r'})]\,\, U + U^{-1}\pp_2 U
\eqno{(20)}
$$
On the other hand, one may rewrite (14) in terms of the transformed quantities:
$$\eqalign{&\pp_1 {A'}_{\!\! 2}^a-\pp_2 {A'}_{\!\! 1}^a +f_{bc}\,^a
{A'}^b_{\!\! 1} {A'}^c_{\!\! 2} -{1\over\kappa}{\textstyle
\sum}{I'}^a_{\!\!\N1}
\delta \left({\bf r}-{\bf q_\N1}\right) = 0\cr
&I'_k\equiv g'_kK_k {g'}^{-1}_k\cr}
\eqno{(21)}
$$
Eq.~(21) determines an expression for $A'_2$ that apparently differs from (20);
of course
the two must agree.

To proceed, take $U$ to satisfy
the following relation
$$
A'_1 = U^{-1}A_1 U + U^{-1}\pp_1 U=0
\eqno{(22)}
$$
{}From Eq.~(21), we have
$$
A'_{2}= {1\over\kappa\pp_1}{\textstyle \sum}I'_{\N1}
\delta \left({\bf r}-{\bf q_\N1}\right) ={1\over\kappa}{\textstyle \sum}
G\left({\bf r}-{\bf q_\N1}\right)I'_{\N1}
\eqno{(23)}
$$
where the operator ${1\over \pp_1}$ is represented by $\int d^2\! r'G({\bf
r}-{\bf r'})$.
Inserting this into (19), we finally get the Darboux transformed Lagrangian,
$$\eqalign{L&= \sum_k[{\bp'}_{\!\!\N1}\cdot {\bf{\dot q}}_\N1 +
2\tr K_\N1 {g'}_{\! \N1}^{ - 1}{\dot {g}}'_\N1 ]\cr
&-\sum_k{1\over 2M_\N1} \left[({p'}^{x}_{\!\!\N1})^2+({p'}^{y}_{\!\!\N1}
+{1\over\kappa}\sum_l G({\bf q}_\N1^{\ }-{\bf q}_\M1^{\ }) {I'}^a_{\N1}
I'_{\M1 a} )^2\right]
 \cr
&-{\kappa\over 3}\int d^2\! r\epsilon^{\alpha\beta\gamma}\tr U^{-1}
\pp_\alpha UU^{-1} \pp_\beta UU^{-1} \pp_\gamma U \cr}
\eqno{(24)}
$$
The last term in (24) is a topological surface term, involving variables that
 decouple from the rest of the Lagrangian, so
we drop it, and the Darboux transformed Lagrangian is simply
$$\eqalign{L&= \sum_k[{\bp}_\N1\cdot{\dot \bq}_\N1 +
2\tr K_\N1 g_{\! \N1}^{\! -\! 1}{\dot {g}}_\N1 ]\cr
&-\sum_k{1\over 2M_\N1} \left[({p}^{x}_\N1)^2+({p}^{y}_\N1
+{1\over\kappa}\sum_l G({\bf q}_\N1^{\ }-{\bf q}_\M1^{\ }) {I}^a_{\N1}
I_{\M1 a} )^2\right]
\cr}
\eqno{(25)}
$$
where primes have been suppressed.
  In order to quantize the Lagrangian in (25), we must still put $2\tr
{\textstyle \sum}K_\N1 g_{\! \N1}^{\! -\! 1}{\dot {g}}_\N1$ into canonical
form.
However, as we show in the Appendix, one may obtain the required commutation
relations [Eq.~(28.b) below] without the complicated passage to Darboux
variables.
So we remain with (25), but a few comments are in order. As mentioned earlier,
special care should be taken in choosing $U$ [as a solution to (22)]
so that (20) be consistent with (23).
As we show in Section~IV, the consistency of the Schr\"odinger equation puts
a restriction on $G$. Our consistent choice is
$$
G({\bf r})={1\over 2}\delta (y)\epsilon (x)
\eqno{(26)}
$$
The self interaction can be ignored; this is
equivalent to setting $G (0)\equiv 0$.
Hence, we conclude that an accurate expression of
the connection for the $k$-th variable is
$${\bf {\cal  G}}_k={1\over\kappa}
\sum_{\M1\not=\N1} {\bf G}({\bf q}_\N1^{\ }-{\bf q}_\M1^{\ }) {I}^a_{\N1}
I_{\M1 a}
\eqno{(27)}$$
with $G_x=0$ and $G_y(\br)=G(\br)$.

Now, it is clear that the fundamental commutation relations are
$$[{q}^i_\N1 ,{ p}^j_\M1 ]=i\delta_{\N1\M1}\delta^{ij} \eqno{(28.a)}
$$
$$
[ I_{\N1 a},{ I}_{\M1 b}]=-if_{ab}\,^c{I}_{\N1 c}\delta_{\N1\M1} \eqno{(28.b)}
$$
and the Hamiltonian of the system is
$$
H=\sum_k{1\over 2M_\N1} \left[{\bp}_\N1-{1\over\kappa}
\sum_{\M1\not= \N1}{\bf G}({\bf q}_\N1-
{\bf q}_\M1) {I}^a_{\N1}
{I}_{\M1 a} \right]^2
\eqno{(29)}$$
[Eqs.~(28.b) are derived in the Appendix.] The
 Hilbert space may be realized in a wave--function representation,
$\phi_{m_1,\dots,m_N}(\bq_1,\dots,\bq_N) $ and each operator is realized by
$${\bf q}_\N1\phi_{m_1,\dots,m_\N1 ,\dots,m_N}=
{\bf q}_\N1\phi_{m_1,\dots,m_\N1 ,\dots,m_N}
\ \ \ \ \ \ \
\eqno{(30.a)}$$
$${\bf p}_\N1\phi_{m_1,\dots,m_\N1 ,\dots,m_N}={\pp\over
i\pp{\bf q}_\N1}\phi_{m_1,\dots,m_\N1 ,\dots,m_N}
\ \ \ \ \ \eqno{(30.b)}$$
$$({I}^a_\N1\phi)_{m_1,\dots,m_\N1 ,\dots,m_N}=({I}^a_\N1 )_{m_\N1^{\ } m'_\N1}
\phi_{m_1,\dots,m'_\N1 ,\dots,m_N}
\eqno{(30.c)}$$
where $({I}^a_\N1 )_{m_\N1^{\ } m'_\N1}$ is a corresponding matrix
representation of
the commutation relations (28.b) ({\it i.e.} a Lie group algebra).
Hence the Schr\"odinger equation is simply
$$\eqalign{
i\pp_t \phi (&\bq_1,\dots,\bq_N)\cr =&
-\sum_k{1\over 2M_\N1} \left(
\nabla_k
-{i\over\kappa}\sum_{\M1\not= \N1}{\bf G}({\bf q}_\N1^{\ }-{\bf q}_\M1^{\ })
{I}^a_{\N1}
{ I}_{\M1 a}^{\ }\right)^2\phi (\bq_1,\dots,\bq_N)\cr}
\eqno{(31)}$$

In case the particles are identical, the exchange symmetry  should be imposed
on the wave--function.
Namely, we require the wave--function to satisfy
$$\eqalign{&
\phi_{m_1,\dots,m_\M1 ,\dots,m_\N1 ,\dots,m_N}(\bq_1,\dots,
\bq_\M1 ,\dots,\bq_\N1 ,\dots,\bq_N)
\cr =&\pm\phi_{m_1,\dots,m_\N1 ,\dots,m_\M1 ,\dots,m_N}(\bq_1,\dots,\bq_\N1
,\dots,\bq_\M1
,\dots,\bq_N)
\cr}
\eqno{(32)}$$
for the boson/fermionic cases.

\goodbreak\bigskip
\line{\bf III.~~A field theoretic description of the
non--Abelian Chern--Simons particles  \hfil}
\nobreak\medskip\nobreak\noindent

Motivated by the fact that the Abelian Schr\"odinger field describes
Abelian charged particles when second quantized,
one may guess that a non--Abelian second quantized Schr\"odinger field
describes
particles carrying  non--Abelian charges.
Since we are considering such particles in  interaction
with a non-Abelian gauge field, whose dynamics is governed by the
Chern--Simons term, the Lagrangian for the system should
contain the Schr\"odinger Lagrangian
and the Chern-Simons term with a minimal coupling between them.

Thus, consider the Lagrange density,
$$
{\cal L} = i\psi^\dagger D_t \psi -{1\over 2}({\bf D}\psi)^\dagger\cdot ({\bf
D}\psi)
-\kappa \epsilon^{\alpha\beta\gamma}
\tr (
A_\alpha \pp_\beta A_\gamma +{2\over 3}A_\alpha A_\beta A_\gamma)
\eqno{(33)}$$
where the covariant derivative $D_\mu$
 is defined by
$$D_\mu\equiv \pp_\mu + A_\mu
\eqno{(34)}$$
and the mass has been scaled to unity.
Also we assume that the Schr\"odinger fields  is in a certain
representation of the group generator, $T^a$.
Let us record here the Euler--Lagrange equations.
$$iD_t \psi= -{1\over 2}{\bf D}\cdot{\bf D}\psi
\eqno{(35)}$$
$$
{\kappa \over 2} \epsilon^{\alpha\beta\gamma}F_{\beta\gamma} = J^\alpha
\eqno{(36)}$$
The non--Abelian charge density and spatial current density read
$$\eqalign{  &J^0\equiv \rho = T_a \rho^a=-i T_a(\psi^\dagger T^a\psi) \cr
&{\bf J}=-{1\over 2} T_a [\psi^\dagger T^a{\bf D} \psi-({\bf D}\psi)^\dagger
T^a\psi] \cr}
\eqno{(37)}$$
which satisfy a covariant continuity equation (as a consequence of the
Euler--Lagrange
equation),
$$
\pp_\mu J^\mu+[A_\mu,J^\mu]=0
\eqno{(38)}$$

Since the physical system in (33) is certainly constrained,
we follow the symplectic methods of Hamiltonian reduction$^{16}$ to find the
required
symplectic structure. For this purpose, we rewrite the Lagrange
density (33) in canonical Darboux form with constraint,
$$
{\cal L} = i\psi^\dagger {\dot \psi}-2\kappa\tr {\dot A}_1A_2
 -{1\over 2}({\bf D}\psi)^\dagger\cdot ({\bf D}\psi)
+2 \tr A_0(\kappa B +\rho )
\eqno{(39)}$$
where irrelevant total derivative terms are dropped. It is clear from
(39) that the Lagrange multiplier $A_0$, enforce the constraint
$$
B+{\rho\over \kappa}=0
\eqno{(40)}$$
It should be noted that Eq.~(40) is of the same  form as Eq.~(14). Using
definitions in (16), the constraint may  again be solved by the
relation (15). Inserting this
solution into (39), we are left with
$$\eqalign{
{\cal L} &= i\psi^\dagger {\dot\psi}+\kappa {\dot A}_{1a}
\int d^2\! r' {\cal M}^a\,_{b}({\bf r},{\bf r'})[\pp'_2 A_1^b({\bf r'})+
{1\over \kappa}\rho^b({\bf r'})]\cr
-&{1\over 2}\{ |(\pp_1 +A_1)\psi|^2 +|(\pp_2 +
\int d^2\! r'T_a {\cal M}^a\,_{b}({\bf r},{\bf r'})[\pp'_2 A_1^b({\bf r'})+
{1\over \kappa}\rho^b({\bf r'})])\psi|^2\}\cr}
\eqno{(41)}$$

While the constraint has been eliminated, the 1--form in ${\cal L}dt$ is not
canonical.
To effect the  Darboux transformation to canonical variables, let us
rewrite (41) in terms of the transformed, primed, variables.
$$\eqalign{&\psi' = U^{-1}\psi,\ \ \ A'_1 = U^{-1}A_1 U +U^{-1}\pp_1 U \cr
&A'_2 = U^{-1}\int\!\! d^2\! r'T_a {\cal M}^a\,_{b}({\bf r},{\bf r'})
[\pp'_2 A_1^b({\bf r'})\!+\!
{1\over \kappa}\rho^b({\bf r'})]
\,\, U +U^{-1}\pp_2 U \cr}
\eqno{(42)}$$
Then, the resulting Lagrangian is
$$\eqalign{
{\cal L} &= i{\psi'}^\dagger {\dot\psi}'+\kappa ( {{\dot A}'^a}_{ 1}){A'}_{2a}
-{1\over 2}{\bf |}\,(\nabla -{\bf A}')\psi'{\bf |}^2 \cr
-&{\kappa\over 3}\int d^2\! r \epsilon^{\alpha\beta\gamma}
\tr U^{-1} \pp_\alpha UU^{-1} \pp_\beta UU^{-1} \pp_\gamma U \cr}
\eqno{(43)}$$
Here, we have used (40) and dropped unnecessary boundary terms for simplicity.
Since (40) is transformed covariantly by (42),
$A'_1$ and $A'_2$ satisfy the transform of relation (40),
$$
\pp_1 A'_2-\pp_2 A'_1+[A'_1,A'_2]-{\rho'\over \kappa}=0
\eqno{(44)}$$
where $\rho'$ denotes $-iT_a{\psi'}^\dagger T^a \psi'$.
 Now, as in Section~II, take $U$ to satisfy
$$
A'_1=U^{-1}A_1 U + U^{-1}\pp_1 U=0
\eqno{(45)}$$
Although Eq.~(42) with $U$ solving (45), gives an
explicit expression for $A'_2$, it is more convenient to obtain the expression
for $A'_2$
from (44) (the $U$ should be
consistently chosen): the solution of (44) with $A'_1=0$ is simply
$$
A'_2={1\over \kappa\pp_1} \rho'
\eqno{(46)}$$
Inserting (45) and (46) into (43), the desired  Lagrange
density is finally given by
$$
{\cal L} = i\psi^\dagger {\dot\psi}
-{1\over 2}\{ |\pp_1\psi|^2 +|(\pp_2 +
{1\over \kappa\pp_1} \rho)\psi|^2\}
\eqno{(47)}$$
where the prime is dropped and the decoupled topological degree of
freedom [$\tr(U^{-1}d U)^3$] is omitted.

Upon quantization, the Lagrange density (47) provides a (bosonic)
equal--time commutation relation.
$$
[\psi_n({\bf r}),\psi_m^\dagger ({\bf r'})] =
\delta_{nm}\delta ({\bf r}\!-\!{\bf r'})
\eqno{(48)}$$
(For definiteness and simplicity, we take a bosonic algebra; a similar
analysis can be easily given with a fermionic degree of freedom.) Then
the Hamiltonian density for the system is given by
$$
{\cal H}=
{1\over 2}\{ |\pp_1\psi|^2 +|[\pp_2 +
{1\over \kappa}\int d^2r' G({\bf r}\!-\!{\bf r'}) \rho({\bf r'})]\psi|^2\}
\eqno{(49)}$$
where ${1\over \pp_1}$ is represented by $\int\! d^2 r' G(\br-\brp )$ as in
Section~II.

To describe the Hamiltonian accurately, we begin by defining operator
covariant derivative
$$
{\bf D}\psi=(\nabla - {\bf A})\psi
\eqno{(50)}$$
where ${\bf A}$ denotes
$$
{\bf A}={1\over \kappa}\int d^2r' {\bf G}({\bf r}\!-\!{\bf r'}) \rho({\bf r'})\
\
\eqno{(51)}$$
{}From the commutation relation, it follows that
$$
[A_{ia}({\bf r})(T^a)_{mn},\psi_n({\bf r'})] ={i\over\kappa}
G_i ({\bf r}\!-\!{\bf r'})(T^aT_a\psi({\bf r'}))_m
\eqno{(52)}$$
Note that ${\bf G} ({\bf r})$  is ill defined at the origin. If we prescribe
that
${\bf G}$ vanishes
at the origin, $\psi({\bf r})$ and $\psi^\dagger ({\bf r})$  commute with ${\bf
A}(\br)$ and
so no ordering problem arises.
It can be easily checked that also no ordering problem arises in
relation with ${\bf A}\cdot{\bf A}$.
With this prescription, the Hamiltonian for the system is
\def\bD{{\bf D}}
\def\bG{{\bf G}}
$$
H={1\over 2}\int d^2r (\bD\psi)^\dagger\cdot(\bD\psi)
\eqno{(53)}$$
The operator field equation of motion follows by commutation with $H$.
$$\eqalign{i{\dot\psi} (\br)&=[\psi (\br), H] \cr
 &=-{1\over 2}\bD\cdot\bD\psi (\br)-iA_0\psi (\br)\cr
 &+{1\over 2 \kappa^2}\int d^2r'\bG ({\bf r'}\!-\!{\bf r})\cdot\bG ({\bf
r'}\!-\!{\bf r})
 \psi^\dagger (\brp)T_aT_b\psi (\brp)\,T^aT^b\psi (\br)\cr
&\equiv \hat O \psi \cr}
\eqno{(54)}$$
$A_0$ in (54) is given by
$$
A_0(\br)=-{1\over \kappa}\int d^2r'\bG ({\bf r'}\!-\!{\bf r})\cdot{\bf J}(\br)
\eqno{(55)}$$
where ${\bf J}$ is the non-Abelian current-density operator, which takes the
same
form as (37). [Eqs.~(51)and (55) solve
(36) when the covariant
continuity equation (38) is used and ordering issues are ignored.]
The last term in the second equality
of (54) is a quantum correction arising from reordering.

Note that the number density operator $\rho_N\equiv \psi^\dagger\psi$ satisfies
the usual
continuity equation:
$$
{\dot\rho}^{\ }_N = i[H,\rho^{\ }_N]= -\nabla\cdot{\bf j}_N
\eqno{(56)}$$
where ${\bf j}_N$ is the U(1) spatial current density:
$$ {\bf j}_N={1\over 2i}[\psi^\dagger \bD\psi-(\bD\psi)^\dagger\psi]
\eqno{(57)}$$
Thus, the number operator $N=\int d^2r \rho^{\ }_N(\br)$ commutes with
the Hamiltonian and also satisfies the  algebra
$$[N,\psi_m(\br)]=-\psi_m(\br),\ \ [N,\psi^\dagger_m(\br)]=\psi^\dagger_m(\br)
\eqno{(58)}$$

Now it is a simple matter to construct the N--particle state.
The vacuum state $|0>$  is annihilated by $\psi_m(\br)$:
$$\psi(\br)|0>=<0|\psi^\dagger(\br)=0
\eqno{(59)}$$
and, therefore, it is a
zero-eigenvalue eigenstate for the both  N and H.

We define the $N$--particle state by $N$ successive operations of $\psi_m(\br)$
 on the vacuum
bra $<0|$.  In this way, we are led to the $N$--particle wave function,
$$\phi_{m_1,\dots,m_N}({\bf r}_1,\dots,{\bf r}_N)=<0|\psi_{m_1}({\bf
r}_1),\dots,
\psi_{m_N}({\bf r}_N)|\Phi>
\eqno{(60)}$$
where $|\Phi>$  is a general state and the $N$--particle amplitude is selected
by
projecting $|\Phi>$  onto the $N$--particle state.

When one computes the time evolution of the Schr\"odinger wave function
in (60) for $N=1$, one simply gets
$$
i\pp_t\phi_m(\br)=<0|(\hat O \psi)_m|\Phi> =-{1\over 2}\nabla^2 \phi_m(\br)
\eqno{(61)}$$
where (54) and (59) are  used to get the second equality:
specifically, one needs
$$
<0|(\hat O \psi)_m =-{1\over 2}\nabla^2<0| \psi_m
\eqno{(62)}$$
Hence the one--particle problem is free---there are no self
interactions.

For the two--particle Schr\"odinger equation, we begin with
$$\eqalign{
i&\pp_t\phi_{m_1m_2}({\bf r}_1,{\bf r}_2)=<0|(\hat O \psi)_{m_1}({\bf r}_1)
\psi_{m_2}({\bf r}_2)+\psi_{m_1}({\bf r}_1)(\hat O
\psi)_{m_2}({\bf r}_2)|\Phi> \cr
&=<0|(\hat O \psi)_{m_1}({\bf r}_1)
\psi_{m_2}({\bf r}_2)+(\hat O \psi)_{m_2}({\bf r}_2)
\psi_{m_1}({\bf r}_1)+ [\psi_{m_1}({\bf r}_1),(\hat O
\psi)_{m_2}({\bf r}_2)]|\Phi> \cr}
\eqno{(63)}$$
Upon using (62) and computing of the commutator
$[\psi_{m_1}({\bf r}_1),(\hat O
\psi({\bf r}_2))_{m_2}]$, one
finds the two particle Schr\"odinger equation to be
$$
i\pp_t\phi ({\bf r}_1,{\bf r}_2)=-{1\over 2}\left[
(\nabla_1\!+\!{i\over\kappa}{\bf G}({\bf r}_1\!-\!{\bf r}_2)T^a_1T_{2a})^2
+(\nabla_2\!+\!{i\over\kappa}{\bf G}({\bf r}_2\!-\!{\bf
r}_1)T^a_2T_{1a})^2\right]
\phi ({\bf r}_1,{\bf r}_2)
\eqno{(64)}$$
where the operator ${T}^a_\N1\phi$  is defined by
$$({ T}^a_\N1\phi)_{m_1,\dots,m_\N1 ,\dots,m_N}=({T}^a_\N1)_{m_\N1^{\ } m'_\N1}
\phi_{m_1,\dots,m'_\N1,\dots,m_N}
\eqno{(65)}$$
By a similiar straightforward computation, the  $N$--body Schr\"odinger
equation is
$$\eqalign{
i\pp_t\phi({\bf r}_1,\dots,{\bf r}_N)&=H_N\phi({\bf r}_1,\dots,{\bf r}_N)\cr
&=-{1\over 2}\sum_\N1 [
\nabla_\N1+{i\over\kappa}\sum_{\M1\not= \N1}{\bf G}({\bf r}_\N1-{\bf r}_\M1)
T^a_\N1T_{\M1 a}]^2
\phi({\bf r}_1,\dots,{\bf r}_N)\cr}
\eqno{(66)}$$
which coincides with (31), for identical bosons with their common
mass scaled to unity ($T^a_k$ corresponds to $iI^a_k$).

\goodbreak\bigskip
\line{\bf IV.~~Interpretation of the connection ${\cal G}_k$ \hfil}
\nobreak\medskip\nobreak\noindent

The connection  ${\cal G}_k$ (27) involves the delta function (26); in the
Schr\"odinger equation (31) or (66) it occurs squared, which requires
well--definition. It is known how to deal
with this problem.$^{18}$
First, regularize the delta function in terms of a peaked function of  width
$\l$ and
 height $o({1\over l})$. Then, equipped with this regulated delta
function, solve the Schr\"odinger equation within a small region
where the connection does not vanish. Finally by a limiting
procedure, which takes $l$  to zero, one arrives at a boundary
condition for the wave function (replacing the set of
points on which the connection is singular). Once the above procedure
is implemented, one finds that the 2--body Schr\"odinger
equation is equivalent to
$$
i\pp_t\phi ({\bf r}_1,{\bf r}_2)=-{1\over 2}(\nabla^2_1+\nabla^2_2)
\phi ({\bf r}_1,{\bf r}_2)
\eqno{(67)}$$
for $y_1\not= y_2$  with a cut at $y_1=y_2\equiv y$. On the cut the boundary
condition
is
$$\eqalign{
&\phi \Bigl({\bf r}_1=(x_1,y+0^+),{\bf r}_2=(x_2,y)\Bigr)\cr
&=e^{{i\over 2 \kappa}{\rm \epsilon}(x_1-x_2)T^a_1
T_{2a}}\phi \Bigl({\bf r}_1=(x_1,y+0^-),{\bf r}_2=(x_2,y)\Bigr)\cr}
\eqno{(68)}$$
The $N$--body  equation becomes
$$
i\pp_t\phi ({\bf r}_1,\dots,{\bf r}_N)=-{1\over 2}\sum\nabla^2_\N1
\phi ({\bf r}_1,\dots,{\bf r}_N)\ \  {\rm for}\ \  y_k\not=y_l \ (k\not=l)
\eqno{(69)}$$
with boundary conditions on the cut $y_k=y_l\equiv y_{kl}\ (k\not=l)$
$$\eqalign{
&\phi \Bigl({\bf r}_1,\dots,{\bf r}_k=(x_k,y_{kl}+0^+),\dots,
{\bf r}_l=(x_l,y_{kl}),\dots,{\bf r}_N\Bigr)\cr &=
e^{{i\over 2\kappa}{\rm \epsilon}(x_k-x_l)T^a_k
T_{la}}\phi \Bigl({\bf r}_1,\dots,{\bf r}_k=(x_k,y_{kl}+0^-),\dots,
{\bf r}_l=(x_l,y_{kl}),\dots,{\bf r}_N\Bigr)\cr}
\eqno{(70)}$$

As claimed in Section~II, the Green's function $G$ in (23) is restricted by
the consistency of the Schr\"odinger equation. If we
choose, for example, $G$ as
$$G(\br)={1\over 2} \delta (y)(\epsilon (x) +C)\ \ (C\not= 0)
\eqno{(71)}$$
the resulting $2$--body Schr\"odinger equation reads
$$
i\pp_t\phi ({\bf r}_1,{\bf r}_2)=-{1\over 2}\left[
\eqalign{&(\pp_1^x)^2+(\pp_2^x)^2\cr
+&(\pp_1^y-{i\over 2\kappa}\delta(y_1-y_2)(\epsilon (x_1-x_2)+C)T^a_1T_{2a})^2
\cr
+&(\pp_2^y-{i\over 2\kappa}\delta(y_2-y_1)(\epsilon (x_2-x_1)+C)T^a_2T_{1a})^2
\cr}\right]
\phi ({\bf r}_1,{\bf r}_2)
\eqno{(72)}$$
Since $\epsilon (x_1-x_2)+C\not= -(\epsilon (x_2-x_1)+C)$, inconsistent
boundary conditions
are obtained  on the cut. Moreover, for $C\not= 0$ Eq.~(72) is not
Galileo--invariant.
[It should be noted that the boundary condition in (70) is consistent with the
exchange-symmertry of the wave function. An equivalent boundary condition is
postulated
in Ref.~[19].]

In Refs.~[13] and [14], there appears a  Schr\"odinger equation where the
potential is
the complex Knizhnik--Zamolodchikov connection.
$$
i\pp_t\phi'= H'_N \phi'
\eqno{(73)}$$
$$\eqalign{
&H'_N=-{1\over 2}\sum_\N1 [
\nabla_\N1 +{i\over\kappa}\sum_{\M1\not= \N1}
{{\bf G}_{KZ}}({\bf r}_\N1 -{\bf r}_\M1)T^a_\N1 T_{\N1 a}^{\ }]^2 \cr
&{{\bf G}}_{KZ}(\br)={1\over 2\pi(x+iy\,)}\left(\matrix{\!\!\phantom{-\!}i\cr
\! -\!1\cr}
\right)\cr}
\eqno{(74)}$$
[Notation here
is changed from complex coordinates to real $(x,y)$ coordinates.] To show that
the equation in (73) is equivalent to (31) or (66), we must find a
time--independent $V$ that connects these equations.
$$
\phi'=V\phi,\ \ \ H'_N =VH_N V^{-1}
\eqno{(75)}$$
Eq.~(75) holds when $V$ satisfies
$$V^{-1}[
\nabla_\N1 +{i\over\kappa}\sum_{\M1 \not= \N1}{{\bf G}_{KZ}}({\bf r}_\N1 -
{\bf r}_\M1)T^a_\N1 T_{\M1 a}^{\ }]
V ={i\over\kappa}\sum_{\M1 \not= \N1}{\bf G}({\bf r}_\N1 -{\bf r}_\M1)T^a_\N1
T_{\M1 a}^{\ }
\eqno{(76)}$$
For the two particle case, this is solved by
$$\eqalign{&V=e^{{1\over 2\pi\kappa}a(\br_1-\br_2)T^a_1T_{2a}}\cr
&a(\br)=
\ln r -i\tan^{-1}\left|{x\over y}\right|\epsilon (x)\epsilon (y) +i{\pi\over 2}
\cr}
\eqno{(77)}$$
where $\tan^{-1} x$ lies in $[0,{\pi\over 2}]$ for $x\geq 0$.
It should be noted that $V$ is a single--valued
function of ${\bf r}_1$  and ${\bf r}_2$ and multi--valuedness does not arise.
In addition, the transformation respects the exchange symmetry of
the wave function since $V(\br_1,\br_2)=V(\br_2,\br_1)$.
Solving (76) for general $N$, remains an open problem.

We have not here included a non--Abelian point interaction, which presumably
would protect
the conformal invariance of the model in field theoretical perturbation theory,
just
as it does in the Abelian model.$^{11}$ This topic is under further
investigation.$^{20}$
\def\ttt{{\theta^a}}
\def\A1{i}
\def\B1{l}
\def\C1{j}
\vfill\eject
\centerline{\bf APPENDIX}
The  purpose of this Appendix is to provide  a  symplectic structure
for the non--Abelian charge $I$. While this problem is analyzed
in many places,$^{21}$the
group is restricted to $SO(n),\  SU(n)$ or especially
$SU(2)$, and the discussion is quite involved since explicit Darboux variables
are
constructed before canonical commutators are found. Here we
give a simple derivation for a general group without passing to
canonical Darboux variables. (The only required
condition is that the Lie algebra possesses a nonsingular  metric.)
As stated in Section~II, the symplectic structure for the non--Abelian charge
$I$
is given by $2\tr Kg^{-1}\dot g$, where $I=gKg^{-1}$ and $K$ is
time independent.

Let us take part of (7) to describe the dynamics of $I$.
Namely, consider
the Lagrangian
$$L_I =2\tr Kg^{-1}\dot g+2\tr I A
\eqno{(A.1)}$$
For this portion of the complete problem, $A$ is viewed as ``external''.
When the Lagrangian  is varied with respect to $g$, the Euler--Lagrange
equation
is
$$\dot I + [ A,I] = 0
\eqno{(A.2)}$$

To find the  classical phase space, parametrize the group element $g$ with
 coordinates $\theta^a$ where index $a$ ranges over the dimension of the group,
which coincides with the  number of generators. The Lagrangian $L_I$
may be rewritten in terms of $\theta^a$.
$$L_I =a_a(\theta){\dot \theta}^a-I^aA_{a}
\eqno{(A.3)}$$
It is convenient to express $a_a(\theta)$ as
$$
a_a=-C_a\,^bI_b
\eqno{(A.4)}$$
where the invertible matrix $C_a\,^b$ is defined by the relation;
$C_a\,^bT_b=\pp_ag
g^{-1}$.
The symplectic two form $\omega_{ab}\equiv \pp_a a_b -\pp_b a_a$ may be
computed using (A.4) and the definition $C_a\,^b$ to give
$$
\omega_{ab}=C_a\,^cC_b\,^d I^e f_{cde}
\eqno{(A.5)}$$
which is not invertible since there are zero modes $p_{i}^a$.
$$
p^a_{i}\omega_{ab}=0
\eqno{(A.6)}$$
[In fact we do not meet the inverse for $\omega_{ab}$ in out final formula
(A.13), but
it arises at intermediate steps in our derivation.]

If we use a projection operator $P_a\,^b$  that satisfies
$$
P_a\,^bP_b\,^c=P_a\,^c,\ \ \ \  P_a\,^b\omega_{bc}=0
\eqno{(A.7)}$$ whose rank coincides with the number of zero modes,
it is possible to find an inverse of $\omega_{ab}$
in the projected subspace. Namely, the inverse ${\omega}^{ab}$
is uniquely defined by the
relations
$$\eqalign{&\omega_{ab}{\omega}^{bc}=\delta_a\,^c-P_a\,^c\cr
&{\omega}^{ab}=-{\omega}^{ba},\ \ {\omega}^{ab}P_b\,^c=0\cr}
\eqno{(A.8)}$$

Once ${\omega}^{ab}$ is constructed, we give the fundamental Poisson bracket
between functions $W_l$ of the $\theta$ as$$
\{W_1,W_2\}=\pp_aW_1{\omega}^{ab}\pp_bW_2
\eqno{(A.9)}$$
It can be easily checked that the Jacobi identity
$$\{W_1,\{W_2,W_3\}\}+\{W_2,\{W_3,W_1\}\}+\{W_3,\{W_1,W_2\}\} = 0
\eqno{(A.10)}$$
is insured when
$$
P_a\,^b\pp_bW_l=0
\eqno{(A.11)}$$
Hence we use the bracket (A.9) only between function $W_l$ of
$\theta^a$ that satisfy (A.11).

We now turn to the problem of calculating the bracket between generators $I$.
First noting that
$$
\pp_mI_a=(C^{-1})_a\,^b\omega_{bm}
\eqno{(A.12)}$$
it follows from (A.7) that (A.11) is satisfied by $I$.
Moreover, using (A.5,7,8,12), we immediately conclude that $I$ satisfies the
expected
bracket algebra.
$$\{I_a,I_b\}=\pp_mI_a{\omega}^{mn}\pp_nI_b=-f_{ab}\,^cI_c
\eqno{(A.13)}$$

In this derivation, explicit expressions of the zero modes and the projection
operator are not used. It may be interesting to obtain them.
To find the zero modes explicitly, first construct all Lie
algebra elements $K^a_\A1 T_a$ that solve
$$f_{abc}K^b_\A1 K^c=0
\eqno{(A.14)}$$
[The number of such Lie algebra elements is greater than or equal to the
 rank of the Lie algebra.]
Then, define $p_i^a$ through relations,
$$
I_\A1^aT_a=gK_\A1 g^{-1},\ \ p^a_\A1 =I^b_\A1(C^{-1})_b\,^a
\eqno{(A.15)}$$
and it is simple matter to show that the $p_i$ satisfy (A.6).
Now let us construct  dual vector ${\bar p}_{\A1 a}$  in the following way.
First define a coordinate transformation from the $\ttt$ coordinates to new
coordinates
$(\xi_1,\dots,\xi_\N1 ,\eta_1,\dots,\eta_{d-\N1})$ (where $k$ and $d$ are
respectively
the number of zero modes and the dimension of the group manifold), such that
$${\pp\ttt\over\pp\xi_\A1}=p^a_\A1
\eqno{(A.16)}$$
Then, ${\bar p}_{\A1}$ is defined by
$${\bar p}_{\A1 a} \equiv {\pp\xi_\A1 \over\pp\ttt}
\eqno{(A.17)}$$
and by construction, they satisfy
$${\bar p}_{\A1 a}p^a_\C1 =\delta_{\A1\C1}
\eqno{(A.18)}$$
Notice ${\bar p}_{ia}p^b_i$ satisfies (A.7) and its rank is $k$. Hence it may
serve
as projection operator in (A.8).

We prove that the Lagrangion (A.1) does not depend on the $\xi_\A1$ variables
except for total derivatives.
To show this, consider an infinitesimal transformation
$$
\delta_\epsilon\ttt =\epsilon (t) p^a
\eqno{(A.19)}$$
Then the variation of the Lagrangian  is
$$
\delta_\epsilon L_I={d\over dt}(\epsilon p^a_\A1 a_a)={d\over dt}(-\epsilon
K^a_\A1 K_a)
=\delta_\epsilon {d\over dt}(-\xi_\A1 K^a_\A1 K_a)
\eqno{(A.20)}$$
To get the second equality in (A.20), we use (A.4) and (A.15).
Noting
$$
\delta_\epsilon (L_I+{\dot \xi}_\A1 K^a_\A1 K_a)=0
\eqno{(A.21)}$$
we conclude that $L_I+{\dot \xi}_\A1 K^a_\A1 K_a$  does not
depend on $\xi_\A1$. Hence, when
we transform to the new
coordinates $({\bf \xi},{\bf\eta})$,
only the ${\bf\eta}$  variables are relevant for the classical phase space and
the
${\bf\xi}$
 variables may be
dropped from the Lagrangian.

Finally we remark that explicit Darboux variables for the $SU(n)$ and $SO(n)$
groups have been found, and especially the case of $SU(2)\sim SO(3)$ is well
known.$^{21}$
Let us record an alternate expression for the $SO(3)$ canonical 1--form, which
also
serves  to quantize the isospin algebra.
$$
L'_I={\bf A}({\bf I})\cdot {\dot{\bf I}}
\eqno{(A.22)}$$
Here ${\bf A}(\br )$ is the potential for the Dirac point monopole of strength
$-|g|$
and ${\bf I}$
is restricted to the surface of a sphere of radius $|g|$. Even though the Dirac
monopole
defines the  singular 2--form, $F_{ij}(\br)= -\epsilon^{ijk}{ r^k\over g^2}$,
this possess an inverse on the tangent space of the sphere, and in this way we
obtain
$\{I^i,I^j\}=-\epsilon^{ijk} I^k$.

\vfill\eject
\centerline{\bf REFERENCES}
\medskip
\item{1.} S. K. Wong, {\it Nuovo Cimento\/} {\bf 65 A} (1970) 689.
\medskip
\item{2.}C. Hagen, {\it Ann. Phys.\/} (NY) {\bf 157} (1984) 342,
{\it Phys. Rev.\/} {\bf D 31} (1985) 2135.
\medskip
\item{3.} J. Leinaas and J. Myrheim, {\it Nuovo Cimento\/} {\bf 37 B} (1977) 1.
\medskip
\item{4.} F. Wilczek, {\it Phys. Rev. Lett.} {\bf 49} (1982) 957.
\medskip
\item{5.}{\it The Quantum Hall Effect\/}, R. Prange and S. Girvin, eds.,
(Springer, Berlin, 1990).
\medskip
\item{6.} R. B. Laughlin, {\it Phys. Rev. Lett.\/} {\bf 60} (1988) 2677.
\medskip
\item{7.} R. Jackiw, {\it Ann. Phys.} (NY) {\bf 201} (1990) 83.
\medskip
\item{8.} R. Jackiw and S.-Y. Pi, {\it Phys. Rev. D\/} {\bf 42} (1990) 3500,
(E)
{\bf 48} (1993) 3929.
\medskip
\item{9.} R. Jackiw and S.-Y. Pi, {\it Phys. Rev. Lett.\/} {\bf 64} (1990)
2669,
(C) {\bf 66} (1991) 2682.
\medskip
\item{10.} R. Jackiw and S.-Y. Pi, {\it Nucl. Phys. B\/} (Proc. Suppl.)
{\bf 33C } (1993) 104.
\medskip
\item{11.} G. Lozano, {\it Phys. Lett. B\/} {\bf 283} (1992) 70;
O. Bergman and G. Lozano, {\it Ann. Phys.} (NY) {\bf 229} (1994)
416; D. Freedman, G. Lozano and
N. Rius, {\it Phys. Rev. D\/} {\bf 49} (1994) 1054; G.
Amelino-Camelina, {\it Phys. Lett. B\/} (in press).
\medskip
\item{12.} G. Dunne, R. Jackiw, S.-Y. Pi and C. A. Trugenberger,
{\it Phys. Rev. D\/} {\bf 43} (1991) 1332, (E) {\bf 45} (1992) 3012; G.
Dunne, {\it Comm. Math. Phys.\/} {\bf 150} (1993) 519.
\medskip
\item{13.} E. Verlinde in {\it Modern Quantum Field Theory\/}, edited by
S. Das {\it et al.}
(World Scientific, Singapore, 1991).
\medskip
\item{14.} T. Lee and P. Oh, {\it Ann. Phys.\/} (NY) (in press),
{\it Phys. Lett. B\/} {\bf 319} (1993) 497.
\medskip
\item{15.} V. Knizhnik and A. Zamolodchikov, {\it Nucl. Phys.\/} {\bf B 247}
(1984) 83.
\medskip
\item{16.} L. Faddeev and R. Jackiw, {\it Phys. Rev. Lett.\/} {\bf 60} (1988)
1692.
\medskip
\item{17.} S. Deser, R. Jackiw and S. Templeton, {\it Phys. Rev. Lett.\/} {\bf
48} (1982) 975,
{\it Ann. Phys.\/} (NY)
 {\bf 140} (1982) 372,
 (E) {\bf 185} (1988) 406.
\medskip
\item{18.} A. Kapustin and P. Pronin, {\it Phys. Lett. B} {\bf 303} (1993) 45.
\medskip
\item{19.} H. Lo and J. Preskill, {\it Califonia Institute of Technology
preprint}
(1993) CALT-68-1867.
\medskip
\item{20.} D. Bak and O. Bergman, in preparation.
\medskip
\item{21.} H. B. Nielsen, D. Rohrlich, {\it Nucl. Phys.\/}
 {\bf B 299} (1988) 471; A. Alekseev, L. Faddeev and S. Shatashvili,
{\it J. Geom. Phys.\/} {\bf 3} (1989) 1;  K. Johnson, {\it Ann. Phys.\/} (NY)
{\bf 192} (1989) 104; A. Balachandran, G. Marmo, B. Skagerstam and A. Stern,
{\it Classical Topology and Quantum States\/} (World Scientific, Singapore,
1991).

\par
\vfill
\end